%% file: main_arxiv.tex
\documentclass[letterpaper,twocolumn,10pt]{article}
\usepackage{usenix2019_v3}

\usepackage{tikz}
\usepackage{amsmath}
\usepackage{xspace}
\usepackage{filecontents}
\usepackage{pifont}
\usepackage[linesnumbered,ruled,vlined]{algorithm2e}
\usepackage{enumitem}

\usepackage{filecontents}

\newcommand{\Sys}[0]{\textit{Zepp}\xspace}

\usepackage{makecell}
\usepackage{amsmath}
\usepackage{amsfonts}
\usepackage{booktabs}

\begin{document}

\date{}

\title{Zepp: Accelerating Distributed MoE Serving under Relaxed Balance Constraints}

\author{
{\rm Chang Chen\textsuperscript{*}}\\
{\rm UCSD} \\
{\rm chc278@ucsd.edu}
\and
{\rm Andrew Yang\textsuperscript{*}}\\
{\rm Stanford University}\\
{\rm azyang@stanford.edu}
\and
{\rm Tiancheng Chen}\\
{\rm ETH Zürich}\\
{\rm tiancheng.chen@inf.ethz.ch}
\and
{\rm Jiangfei Duan}\\
{\rm Independent Researcher}\\
{\rm imjfduan@gmail.com}
\and
{\rm Xinwei Qiang}\\
{\rm UCSD}\\
{\rm x2qiang@ucsd.edu}
\and
{\rm Zhongkai Yu}\\
{\rm UCSD}\\
{\rm zhy055@ucsd.edu}
\and
{\rm Xiang Fang}\\
{\rm UCSD}\\
{\rm x8fang@ucsd.edu}
\and
{\rm Yufei Ding}\\
{\rm UCSD}\\
{\rm yufeiding@ucsd.edu}
}

\maketitle

\begingroup
\renewcommand{\thefootnote}{*}
\footnotetext[1]{These authors contributed equally.}
\endgroup

\begin{abstract}
As Mixture-of-Experts (MoE) models continue to scale, serving them increasingly relies on expert parallelism (EP) across a growing number of devices. Yet skewed expert workloads create imbalance across computation, communication, and memory, making load balancing a central optimization objective in distributed MoE serving.

We observe that \textit{balance is not free}: operations introduced to balance one dimension can themselves be expensive or imbalanced. This motivates us to rethink balance as a constraint rather than an optimization objective. We present \Sys, which directly optimizes the bottleneck communication in distributed MoE serving subject to simplified balance constraints on physical resources, i.e., GPUs and NICs. \Sys progressively optimizes inter-node communication across placement, routing, and execution. It first places expert replicas to reduce token communication under GPU constraints, then reshapes communication flows through \textit{split} and \textit{merge} primitives under NIC constraints, and finally partitions and schedules expert computation to overlap the resulting communication. To adapt to dynamic workloads, \Sys jointly coordinates computation, token communication, and expert-weight movement at each iteration. Together, these designs allow \Sys to pursue the most efficient execution rather than a single-dimension balanced one. We implement \Sys and evaluate it against \textbf{7} state-of-the-art MoE serving systems, achieving up to 6.68$\times$ MoE layer speedup and a geometric mean speedup of 1.86$\times$ over the fastest competing baseline.

\end{abstract}

\input{01_introduction}

\input{02_background}

\input{03_Motivation}

\input{04_Method}

\input{05_Evaluation}

\input{06_Conclusion}

\bibliographystyle{plain}
\bibliography{ref}

\end{document}

%% file: 01_introduction.tex
\section{Introduction}
Mixture-of-Experts (MoE) has emerged as a key architecture for scaling large language models (LLMs), enabling substantial growth in model capacity while activating only a small subset of parameters for each token~\cite{lepikhin2021gshard,shazeer2017outrageously}. This sparse activation makes it possible to scale models to hundreds of billions or even trillions of parameters without proportionally increasing computation~\cite{fedus2022switch}. Consequently, MoE has been increasingly adopted by state-of-the-art LLMs~\cite{kimiteam2026kimik3openfrontier,kimiteam2026kimik2openagentic,yang2025qwen3technicalreport,deepseekai2026deepseekv4, tamoe}, making efficient MoE inference critical to delivering low-latency and high-throughput serving at scale.

\begin{figure}
    \centering
    \includegraphics[width=1\columnwidth]{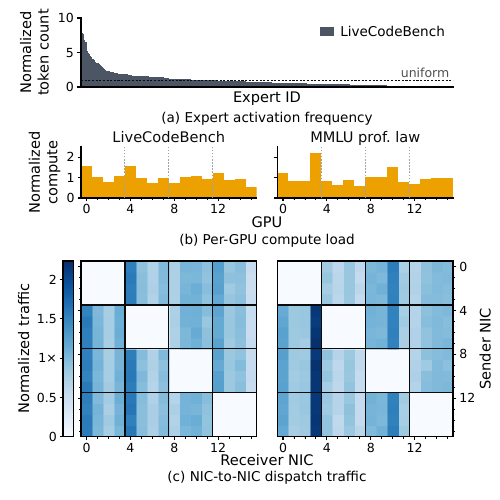}
    \vspace{-30pt}
    \caption{Imbalance captured from realistic MoE traces under two datasets~\cite{yu2026patternschaosforecastingdata}. (a) shows expert activation skew. (b) and (c) show resulting compute and communication imbalance when experts are mapped in round-robin order onto a 4 node, 16 GPU topology with a NIC paired per GPU.}
    \vspace{-20pt}
    \label{fig:intro}
\end{figure}

Serving large MoE models typically relies on expert parallelism (EP), which distributes experts across multiple GPUs and routes each token to the GPUs hosting its selected experts~\cite{lepikhin2021gshard,deepseek2025deepep}. Each MoE layer therefore alternates between token dispatch, expert computation, and token combine, introducing substantial all-to-all communication across GPUs~\cite{lei2026fast,hwang2023tuteladaptivemixtureofexpertsscale}. More importantly, dynamic token-to-expert routing creates highly skewed and time-varying workloads~\cite{Nie_2023,li2023lina,yu2026patternschaosforecastingdata}: as shown in Figure~\ref{fig:intro}, different experts receive different numbers of tokens, leading to uneven expert computation and communication traffic across GPUs and network links. Such imbalance can leave some resources overloaded while others remain idle, making efficient EP execution a central challenge for high-performance MoE serving~\cite{zeng2025efficientmoe,hwang2023tuteladaptivemixtureofexpertsscale,li2023lina}.

Existing systems follow a common principle: better balance leads to better performance. Early approaches balance communication by padding redundant tokens to enforce uniform communication shapes, at the cost of unnecessary data transfer~\cite{zhang2025comet, lepikhin2021gshard, hwang2023tuteladaptivemixtureofexpertsscale}. More recent systems eliminate such redundancy and instead balance traffic across bottleneck resources, e.g., by routing tokens across multiple NICs, but introduce explicit routing overheads~\cite{ren2025fuselink, lei2026fast}. To further balance computation and communication, systems replicate hot experts~\cite{deepseek2025eplb, zhao2026craft,zeng2025efficientmoe} and may periodically remap experts across devices in response to changing loads~\cite{cao2026epic,Nie_2023, moonshot2026moonep}. However, expert remapping, in turn, introduces expensive and inherently imbalanced communication. Reducing the remapping frequency mitigates this overhead, but evenly dispatching tokens across expert replicas to maintain runtime balance can still introduce additional inter-node communication.

These approaches reveal a fundamental balancing dilemma: eliminating one form of imbalance often introduces another source of imbalance or overhead, making balance itself increasingly costly to pursue. Orthogonal techniques, such as communication-computation overlap~\cite{chang2024flux,zhang2025comet,cao2026epic, centauri, Zepplin} and attention-FFN disaggregation~\cite{zhu2025megascaleinfer,wang2025amoe}, can mitigate the performance impact of imbalance, but do not fundamentally resolve this dilemma. More importantly, we observe that better balance does not necessarily translate into better performance, particularly when achieving balance requires additional inter-node communication. This motivates us to rethink the role of balance in distributed MoE serving: rather than treating balance as the optimization objective, we enforce it only as a constraint on physical resources, i.e., GPUs and NICs, and directly minimize the bottleneck inter-node communication.

We present \Sys, a distributed MoE serving system that directly optimizes bottleneck inter-node communication while treating GPU and NIC balance as resource constraints. Rather than pursuing balance in each dimension independently, \Sys progressively avoids, reshapes, and hides inter-node communication.

First, \Sys avoids unnecessary inter-node communication by jointly coordinating expert replication, placement, and token routing under GPU constraints. Replication and placement use historical demand to spread expert replicas across nodes, creating opportunities for locality. Given this placement, locality-preferred routing favors local replicas and consolidates remaining expert requests onto fewer remote nodes. Together, they reduce inter-node traffic without introducing additional communication operations.

Second, \Sys restructures and overlaps the remaining communication through \emph{split} and \emph{merge} under NIC constraints. Unlike conventional overlap designs that primarily split communication, \Sys adjusts communication granularity in both directions: \emph{merge} eliminates redundant inter-node payloads through shared token transfers during dispatch and node-local pre-reduction during combine, while \emph{split} exposes independently schedulable flows and redistributes traffic over NVLink to alleviate bottleneck NICs. \Sys then schedules expert computation around the dependencies of these restructured flows, allowing ready computation and communication to overlap during both dispatch and combine.

Finally, dynamic workload imbalances can cause the historical placement to violate GPU load constraints. Instead of globally remapping experts, \Sys performs intra-node expert swaps that preserve node-level expert coverage and avoid inter-node weight transfers. These weight movements are further coordinated with token communication and expert computation to hide their cost whenever possible.

To conclude, this paper makes three contributions:
\begin{itemize}[leftmargin=*]
    \setlength{\itemsep}{0pt}
    \setlength{\parskip}{0pt}
    \item \textbf{Motivation and analysis.}
    We analyze how balancing expert computation and communication can introduce additional inter-node transfers and movement overhead. This motivates treating expert-replica load balance as a constraint while reducing bottleneck inter-node communication.

    \item \textbf{System design.}
    We design and implement \Sys, which coordinates expert replication, placement, and routing to reduce remote token transfers. Building on these assignments, \textit{split} Xand \textit{merge} restructure the remaining communication, and computation scheduling overlaps the resulting flows. Intra-node expert swaps extend this design to dynamic demand, rebalancing GPU compute load while overlapping expert-weight
    movement.

    \item \textbf{Evaluation.}
    We evaluate \Sys across multiple MoE model architectures, topologies, and input token datasets. Results showcase that \Sys achieves up to $6.68\times$ speedup over existing baselines and up to $2.78\times$ ($1.86\times$ geometric mean) over the fastest competing baseline in end-to-end MoE layer execution.
\end{itemize}
Code is available at \url{github.com/andronius-yang/zepp}.

%% file: 02_background.tex
\begin{figure}[t]
    \centering
    \includegraphics[width=1\columnwidth]{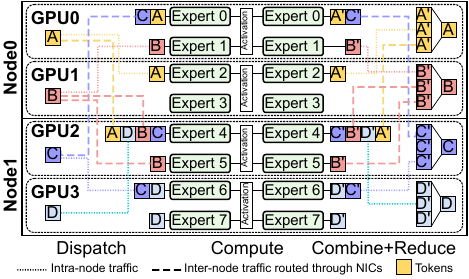}
    \vspace{-25pt}
    \caption{Example of MoE Layer on two nodes, each with two GPUs.}
    \vspace{-10pt}
    \label{fig:moe}
\end{figure}

\begin{figure*}[t]
    \centering
    \includegraphics[width=1\textwidth]{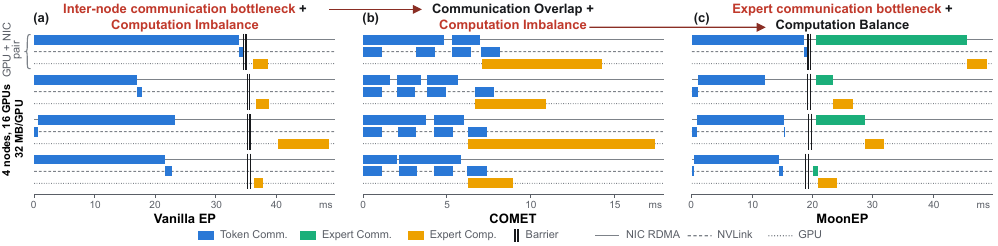}
    \vspace{-25pt}
    \caption{Resource usage timelines for MoE dispatch. Optimizing one resource can expose or introduce a bottleneck in another. (a) Vanilla EP: communication bottleneck and computation imbalance~\cite{lei2026fast,deepseek2025deepep,li2026swiftep,ren2025fuselink}. (b) Communication overlap: computation imbalance remains~\cite{zhang2025comet,aimuyo2025flashmoe,hwang2023tuteladaptivemixtureofexpertsscale,liang2026streamep}. (c) Computation balancing: expert communication becomes the new bottleneck~\cite{moonshot2026moonep,deepseek2025eplb,cao2026epic,zhao2026craft}.}
    \vspace{-10pt}
    \label{fig:imbalance}
    
\end{figure*}

\section{Background}
Mixture-of-Experts (MoE) has become an increasingly prominent architecture in modern large language models. Unlike dense models that activate the same parameters for every token, MoE models sparsely activate only a subset of parameters organized into \textit{experts}. Consequently, increasing model capacity does not necessarily require a proportional increase in per-token computation, enabling substantially larger parameter counts~\cite{shazeer2017outrageously,fedus2022switch,dai2024deepseekmoe}. Recent models illustrate this: Kimi K3 scales to 2.8 trillion total parameters while activating only 104 billion parameters per token~\cite{kimiteam2026kimik3openfrontier}.

To determine the set of activated experts, a learned router dynamically selects the top-$k$ experts to process each \textit{token}, making token-to-expert routing an intrinsic part of MoE execution~\cite{shazeer2017outrageously,lepikhin2021gshard}. To scale large MoE layers across multiple GPUs, \textit{expert parallelism} (EP) distributes experts across different GPUs~\cite{rajbhandari2022deepspeedmoe}. Figure~\ref{fig:moe} illustrates the resulting selection and movement of tokens: during \textit{dispatch}, tokens travel to GPUs hosting their selected experts, where computation is performed. The outputs are then returned during \textit{combine} and aggregated at each token's source.

When EP strategies are deployed to multiple nodes, their token traffic is mapped onto a heterogeneous hierarchy of GPU interconnects. Intra-node GPUs communicate through high-bandwidth, low-latency scale-up fabrics such as NVLink and NVSwitch~\cite{nvidia_nvlink}. Inter-node GPUs rely on NICs and scale-out networks with much lower bandwidth and higher communication overhead. These communication paths utilize distinct data-movement resources: CUDA peer-to-peer copies can utilize dedicated copy engines for intra-node data transfers~\cite{cuda_programming_guide}, while NVSHMEM provides fine-grained data transfer operations that utilize NICs to perform inter-node data transfers~\cite{nvshmem}. By utilizing resources separate from the GPU's compute units, these mechanisms create opportunities to overlap communication with computation.

Dynamic routing decisions therefore naturally produce imbalance, whose effects are shaped by the underlying heterogeneous interconnects. Uneven token assignments across experts create both varying communication volumes between GPUs and uneven computation loads on GPUs hosting those experts~\cite{fedus2022switch,li2023lina,cui2023brainstorm}. Figure~\ref{fig:moe} shows both effects in play: Expert 4 receives four tokens and depends on tokens delivered across nodes from remote GPUs, whilst Expert 3 is not activated in this layer at all. As routing layer decisions are generated independently at each MoE layer, these imbalances shift across tokens and layers throughout model execution.

%% file: 03_Motivation.tex
\section{Motivation}
\begin{figure*}[!t]
    \centering
    \includegraphics[width=1\textwidth]{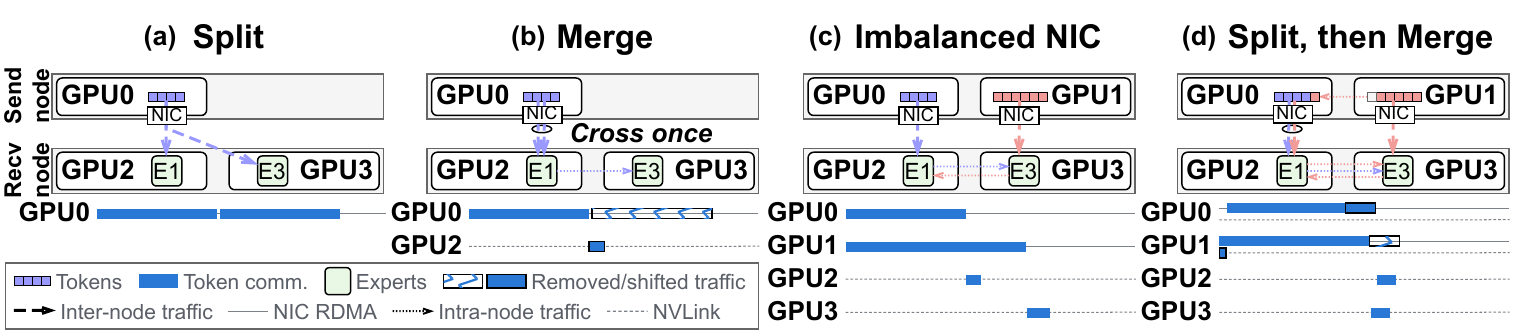}

    \caption{Split and merge during dispatch. (a) Separate expert deliveries repeat a shared input. (b) Merge sends it across nodes once. (c) Unequal source payloads leave NICs imbalanced. (d) Local redistribution and merging spread traffic across NICs.}
    \label{fig:split_merge}
\end{figure*}
Compared with dense distributed serving, MoE serving introduces two unique challenges: load imbalance from dynamic expert activation and communication planning under irregular traffic.
\subsection{Imbalance is Shifted, Not Eliminated}
Existing MoE serving systems increasingly optimize imbalance across different resources. However, we observe that balancing one dimension often shifts the bottleneck to another rather than eliminating it.

\begin{itemize}[leftmargin=*]
    \item \textbf{Communication overlap.} As shown in Figure~\ref{fig:imbalance}(a), vanilla EP suffers from both expensive inter-node token communication and imbalanced expert computation. Existing overlap techniques partition communication and computation to pipeline their execution, effectively hiding part of the communication latency. However, the skewed expert workload remains, leaving the critical path dominated by straggling expert computation (Figure~\ref{fig:imbalance}(b)). Thus, reducing the impact of communication imbalance can expose computation imbalance as the next bottleneck.
    \item \textbf{Computation balancing.} Expert replication and remapping further balance expert computation by redistributing hot experts across GPUs. However, moving expert weights introduces additional inter-node communication, which can become the new bottleneck even after computation is well balanced (Figure~\ref{fig:imbalance}(c)). In other words, eliminating computation imbalance can shift the bottleneck back to communication.

\end{itemize}

These results reveal a recurring pattern: optimizing balance in one dimension can expose or introduce imbalance in another. Rather than pursuing balance itself, we argue that MoE serving should directly optimize the end-to-end critical path, dominated by inter-node communication as shown in Figure~\ref{fig:imbalance}(a), while treating resource balance as a constraint.

\subsection{Split, or Merge?}
Dense-model overlap typically splits large communication operations to enable pipelining. However, MoE's sparse and irregular communication makes finer granularity not always beneficial.
\begin{itemize}[leftmargin=*]
\item \textbf{Redundant Tokens.} In dense communication, different chunks typically carry distinct data, so \textbf{\textit{splitting}} primarily changes granularity without duplicating payloads. In MoE, however, multiple experts selected by the same token may reside on the same remote node. Sending their requests separately repeatedly transfers the same token. As shown in Figure~\ref{fig:split_merge}(a--b), \textbf{\textit{merging}} such transfers allows the shared token to cross the network only once.

\item \textbf{Imbalanced NIC Traffic.} Dense collectives typically distribute regular traffic across participating ranks, allowing fine-grained splitting to expose overlap without substantially changing traffic balance. MoE routing instead produces highly skewed traffic across GPUs and NICs. As shown in Figure~\ref{fig:split_merge}(c--d), \textbf{\textit{splitting}} and locally redistributing traffic can alleviate bottleneck NICs, while subsequent merging removes redundant transfers.

\item \textbf{Message Granularity.} As EP scales to more devices, each GPU serves fewer tokens and individual MoE transfers become smaller, making per-message latency increasingly significant. Unlike dense communication, where large transfers can amortize startup latency, aggressively splitting these already-small MoE messages further increases the number of latency-dominated transfers. \textbf{\textit{Merging}} small transfers therefore becomes necessary to amortize communication latency at scale.

\end{itemize}
While dense-model overlap primarily asks how to \textit{split} communication, MoE serving must decide whether to \textit{split} or \textit{merge}: \textit{split} to alleviate NIC bottlenecks, and \textit{merge} to reduce redundant transfers and communication latency.

%% file: 04_Method.tex
\section{System Design}
\label{sec:system-design}
\begin{figure}[t]
    \centering
    \includegraphics[width=1\columnwidth]{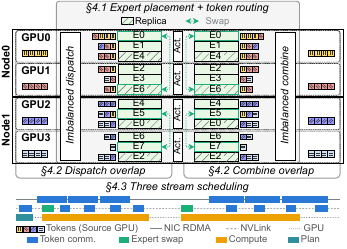}
    
    \caption{Design overview of \Sys. }
    \label{fig:sys}
\end{figure}
\Sys directly optimizes the bottleneck inter-node communication of MoE layers while treating GPU compute and NIC traffic balance as resource constraints. Figure~\ref{fig:sys} summarizes the design. First, under GPU compute load constraints, \Sys jointly coordinates expert replication, placement, and token routing to avoid inter-node transfers whenever local replica capacity permits (\S\ref{sec:placement-routing}). For the remaining transfers, \Sys uses \textit{split} and \textit{merge} to restructure and reduce communication under NIC traffic constraints, and reschedules expert computation around the resulting communication flows to overlap communication with computation (\S\ref{sec:overlap-scheduling}). Finally, as dynamic expert demand shifts GPU compute loads, \Sys uses intra-node expert swaps to restore the load constraints without introducing inter-node weight transfers (\S\ref{sec:expert-swap}). Table~\ref{tab:design-notation} summarizes the notation for expert assignment, GPU compute load, and token communication. 

\begin{table}
\centering
\small
\setlength{\tabcolsep}{3pt}
\renewcommand{\arraystretch}{1.05}
\caption{Design notation. Indices $i,j$ denote GPUs and $e$ an expert.
}
\label{tab:design-notation}
\begin{tabular}{@{}p{0.23\columnwidth}p{0.73\columnwidth}@{}}
\toprule
Symbol & Meaning \\
\midrule
$M^e_{ij}$ & Number of tokens for expert $e$ from GPU $i$ assigned to GPU $j$. \\
$d_i^e$ & Number of tokens for expert $e$ from GPU $i$. \\
$q_j^e$ & Balanced number of tokens for expert $e$ on GPU $j$ after replication; zero if no replica is present. \\
$a_{ij}$ & Inverse bandwidth from GPU $i$ to $j$. \\
$C$ & Relaxation ratio of token budget \\
$R$ & Number of extra expert slots on each GPU \\
\bottomrule
\end{tabular}
\end{table}

\subsection{Expert Placement and Token Routing}
\label{sec:placement-routing}
We first optimize expert placement and token routing, as they reduce inter-node communication without introducing additional communication operations. By changing the number of replicated experts, where expert replicas reside at the beginning, and which replica serves each token at each forward round, \Sys reduces remote token transfers using only the communication already required by MoE execution. We therefore prioritize this optimization before introducing more active communication transformations. These decisions jointly determine GPU compute load and token traffic. \Sys preserves the model's logical expert selections and routing weights while choosing their physical execution locations, without changing the original algorithm.

Considering expert placement and token routing independently can lead to expensive token assignments even when replica loads are balanced. A placement may balance expected compute load across GPUs while clustering popular experts within a few nodes, providing limited node-level expert coverage for local token dispatch. Conversely, evenly distributing each source rank's tokens across expert replicas can introduce unnecessary remote transfers even when local assignments can satisfy the replica load targets. We therefore treat expert-replica load balance as a constraint rather than the routing objective: among assignments satisfying the load constraints, we prefer those with lower communication cost.

\noindent\textbf{Expert replication and placement.}
\Sys replicates experts based on historical demand while prioritizing expert coverage across nodes. In addition to the original slots for evenly distributed expert weights, we reserve $R$ extra slots on each GPU for expert replicas. We greedily assign these slots to the expert with the highest historical token demand per replica, updating its demand per replica after each replication, until all available slots are filled.

\begin{figure}
    \centering
    \includegraphics[width=1\columnwidth]{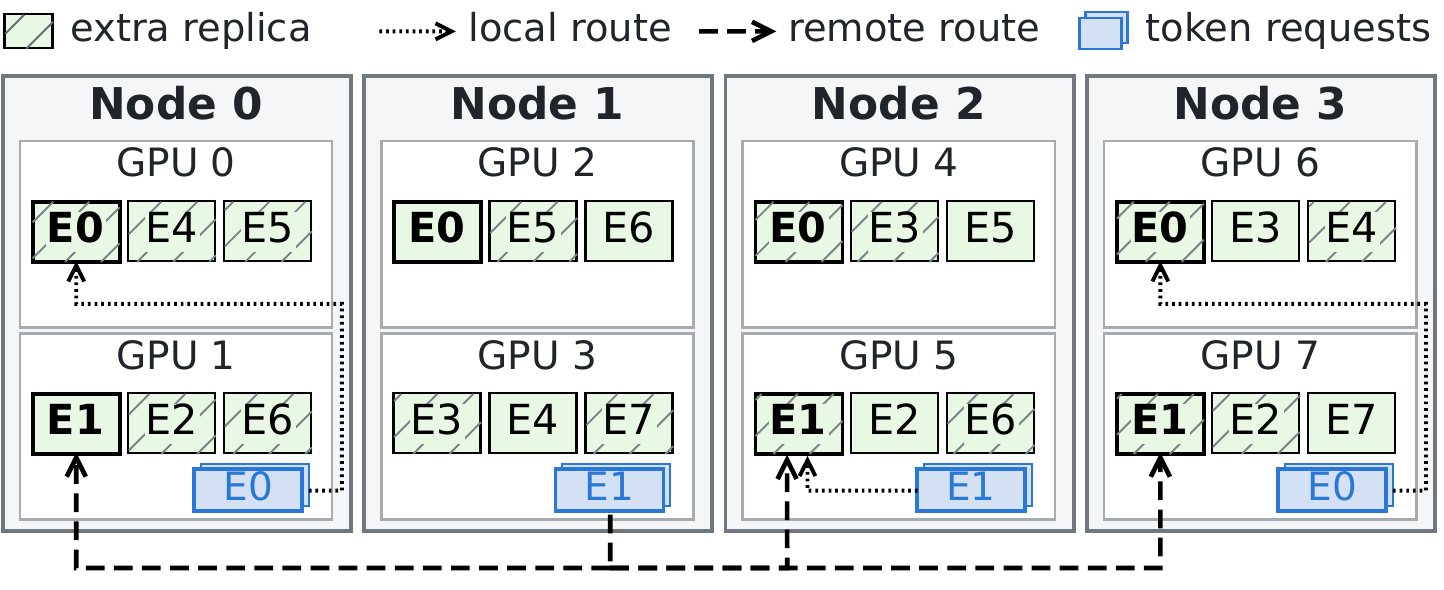}
    \vspace{-25pt}
    \caption{Replica placement and routing showing E0 replicated across all nodes with local token routes. E1 replicated in all but one with inter-node routes for absent node tokens.}
    \label{fig:pl_rt}
\end{figure}

After determining the number of replicas, \Sys places them hierarchically. Following workload priority, we first distribute replicas across nodes in a round-robin manner, and then distribute replicas within each node across GPUs to balance the expected compute load. This hierarchical placement spreads replicas of the same expert across nodes, so that each additional replica expands its node-level coverage rather than creating redundant copies within the same node. The resulting placement is reused across batches, avoiding recurring expert-weight movement while creating locality opportunities for runtime token routing. Figure~\ref{fig:pl_rt} illustrates the resulting heterogeneous expert coverage across nodes. The increased node-level coverage also provides greater flexibility for the intra-node expert adjustments described in \S\ref{sec:expert-swap}.

\begin{figure*}[!t]
    \centering
    \includegraphics[width=1\linewidth]{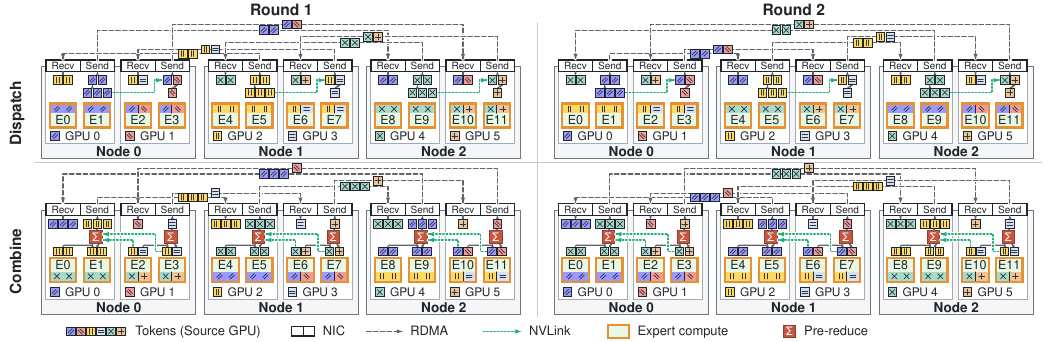}
    \vspace{-25pt}
    \caption{Two rounds of comm.--comp. overlap in dispatch and combine. Tokens inside expert boxes denote current computation; tokens along communication paths belong to other groups being transferred concurrently. Across rounds, received inputs enter expert computation during dispatch, while completed outputs enter pre-reduction ($\Sigma$) and then return transfer during combine.}
    \label{fig:graph}
\end{figure*}

\noindent\textbf{Locality-preferred token routing.}
Given the placement, each rank adaptively adjusts its token dispatch without introducing additional communication operations. For each expert \(e\), a GPU-to-GPU matrix \(M^e\) records the number of tokens assigned from each source GPU to each serving GPU. The row totals are fixed by the batch's expert selections, while \(q_j^e\) denotes the balanced reference load of expert \(e\) on GPU \(j\), computed by evenly dividing the total number of tokens assigned to \(e\) across its replicas, with \(q_j^e=0\) if GPU \(j\) does not host \(e\) and \(\sum_j q_j^e=\sum_i d_i^e\). We formulate communication-minimizing routing under relaxed GPU load constraints as
\begin{equation}
\label{eq:routing-cost}
\begin{aligned}
\min_M\quad & \sum_{e,i,j} a_{ij}M^e_{ij} \\
\text{s.t.}\quad
& \sum_j M^e_{ij}=d_i^e && \forall i,e, \\
& |\sum_i M^e_{ij} - q_j^e| \leq C * q_j^e && \forall j,e, \\
& M^e_{ij}\in\mathbb{Z}_{\geq 0} && \forall i,j,e.
\end{aligned}
\end{equation}
The first constraint preserves the model's original expert selections. The second bounds each replica's load around its balanced reference and, because \(q_j^e=0\) for non-resident experts, prevents routing tokens to GPUs that do not host the selected expert. The second bound also constrains the aggregate token load on each GPU. The relaxation ratio \(C\) therefore controls how far routing may deviate from perfect balance in exchange for lower communication cost. For homogeneous GPUs and equal-size experts, token counts approximate compute load. With high bandwidth of intra-node paths, the objective favors local replicas whenever their available load budget permits. As illustrated by E0 in Figure~\ref{fig:pl_rt}, sufficient replica coverage can avoid inter-node transfers; otherwise, tokens are routed to remote replicas, as illustrated by E1 from Node1. The overhead of solving this optimization problem is modest compared with its performance benefits, as detailed in \S\ref{sec:implementation}.

\subsection{Overlapping Scheduling}
\label{sec:overlap-scheduling}
Although expert placement and locality-preferred token routing reduce inter-node communication volume, the remaining communication can still bottleneck MoE execution. Its performance depends not only on the total traffic volume, but also on how traffic is distributed across the multiple NICs of each node. An overloaded NIC can dominate communication latency even when the aggregate traffic volume is small. We therefore introduce \textit{split} and \textit{merge} operations to restructure communication flows under NIC balance constraints, and schedule expert computation around the resulting communication dependencies to overlap inter-node transfers with computation.

\noindent\textbf{\textit{Split}.}
Conventional EP expresses token dispatch as an all-to-all collective, coupling
many token transfers into a single communication operation. \Sys first
\emph{splits} the all-to-all into independently schedulable send-receive flows
according to their source and destination nodes. This decomposition exposes
the underlying traffic and allows \Sys to explicitly control which NIC carries
each inter-node transfer.
However, the resulting flows can remain highly imbalanced across NICs because
dynamic expert routing generates different amounts of outgoing traffic on
different GPUs. \Sys therefore further \textit{splits} and redistributes these flows
across local GPUs to satisfy NIC load constraints. 

We first analyze the traffic between a pair of nodes to determine how its outgoing load should be distributed across the available NICs. Let $V_i$ denote the
original outgoing traffic from GPU $i$ toward a destination node, and
$x_{ij}$ the amount of this traffic eventually transmitted through the NIC
attached to GPU $j$. When $i=j$, the traffic directly uses its original NIC;
when $i\neq j$, the corresponding payload is first transferred from GPU $i$
to GPU $j$ over NVLink and then sent through GPU $j$'s NIC.

We determine the redistribution by minimizing the additional intra-node data
movement subject to NIC load constraints:
\begin{equation}
\label{eq:nic-split}
\begin{aligned}
\min_{\{x_{ij}\}} \quad
    & \sum_i \sum_{j\neq i} c_{ij}x_{ij} \\
\mathrm{s.t.}\quad
    & \sum_j x_{ij}=V_i, && \forall i,\\
    & \sum_i x_{ij}\le
      (1+C_{\mathrm{NIC}})\frac{V}{G}, && \forall j,\\
    & x_{ij}\ge 0, && \forall i,j,
\end{aligned}
\end{equation}
where $V=\sum_i V_i$, $G$ is the number of NIC-equipped GPUs in the node,
$c_{ij}$ captures the intra-node movement cost from GPU $i$ to GPU $j$, and
$C_{\mathrm{NIC}}$ controls the allowed deviation from perfectly balanced NIC
traffic. Thus, NIC balance is enforced as a constraint rather than the
optimization objective: \Sys keeps traffic on its original NIC whenever
possible and redistributes only the traffic necessary to satisfy the
constraint.

\begin{algorithm}[t]
\caption{Overlapping dispatch and combine.}
\label{alg:overlap}
\small
\DontPrintSemicolon
\SetAlgoNlRelativeSize{-1}
\SetKwFor{After}{after}{}{}
\SetKwFor{ForEach}{for each}{do}{end for}

\KwIn{GPU $(n,g)$; $N$ nodes; $G$ GPUs per node;
      split/merge plan $\mathcal{P}$; routed inputs and
      computation groups}
\KwOut{Combined token outputs $\mathbf{y}$}
$\mathbf{y}\gets\mathbf{0}$\;

{\let\nl\relax \textbf{Dispatch}\;}
\For{$r=1,\ldots,N-1$}{
    $d\gets(n-r)\bmod N;\quad s\gets(n+r)\bmod N$\;
    \ForEach{relay $h$}{
        $\mathrm{Put}_{\mathrm{NVLink}}(\mathrm{pieces}[d,h],(n,h))$\;
    }
    \After{$\mathrm{ready}(\mathrm{relay}[d])$:}{
        $\mathrm{Put}_{\mathrm{RDMA}}(\mathrm{relay}[d],(d,g))$\;
    }
    \After{$\mathrm{ready}(\mathrm{arrival}[s])$:}{
        \For{$\ell=0,\ldots,G-1$}{
            $h\gets(g+1+r+\ell)\bmod G$\;
            $\mathrm{Put}_{\mathrm{NVLink}}(\mathrm{arrival}[s],(n,h))$\;
        }
    }
}
\For{input group $q$ in planned delivery order}{
    \lAfter{$\mathrm{ready}(\mathrm{inputs}(q))$:}{
        \textsc{Compute}($q$)}
}

\BlankLine
{\let\nl\relax \textbf{Combine}\;}
\For{$d=n+1,\ldots,n+N-1,n$ (mod $N$)}{
    $\mathrm{outputs}[d,*]\gets$
        \textsc{Compute}($\mathrm{returnGroup}[d]$)\;
}
\For{$r=1,\ldots,N-1$}{
    $d\gets(n+r)\bmod N;\quad s\gets(n-r)\bmod N$\;
    \After{$\mathrm{ready}(\mathrm{outputs}[d,*])$:}{
        \For{$\ell=0,\ldots,G-1$}{
            $h\gets(g+\ell)\bmod G$\;
            $\mathrm{Put}_{\mathrm{NVLink}}(\mathrm{outputs}[d,h],(n,h))$\;
        }
    }
    \After{$\mathrm{ready}(\mathrm{contributions}[d])$:}{
        $\mathrm{partial}[d]\gets$
            \textsc{PreReduce}($\mathrm{contributions}[d]$)\;
    }
    \After{$\mathrm{ready}(\mathrm{partial}[d])$:}{
        $\mathrm{Put}_{\mathrm{RDMA}}(\mathrm{partial}[d],(d,g))$\;
    }
    \lAfter{$\mathrm{ready}(\mathrm{arrival}[s])$:}{
        \textsc{Reduce}($\mathrm{arrival}[s]$, $\mathbf{y}$)}
}
\textsc{WaitAll}()\;
\Return{$\mathbf{y}$}\;
\end{algorithm}

For homogeneous NICs and $C_{\mathrm{NIC}}=0$, each NIC carries $V/G$ traffic,
and the minimum amount of traffic that must be moved within the node is
\begin{equation}
\label{eq:min-redistribution}
V_{\mathrm{move}}^{\min}
=
\sum_i\left(V_i-\frac{V}{G}\right)^+
=
\frac{1}{2}\sum_i
\left|V_i-\frac{V}{G}\right|.
\end{equation}
As illustrated in Figure~\ref{fig:split_merge}, when GPU1 carries more outgoing
traffic than GPU0, \Sys \textit{splits} only its excess traffic and moves it to GPU0
over NVLink. The selected experts and destination nodes remain unchanged;
only the physical NIC carrying the payload changes. In this way, \Sys trades
the minimum necessary high-bandwidth intra-node movement for reduced traffic
on bottleneck inter-node NICs.

\noindent\textbf{\textit{Merge}.}
NIC-aware splitting exposes flexible physical paths, but can also produce
fine-grained transfers and preserve redundant payloads across independent
flows. \Sys therefore \emph{merges} compatible flows after redistribution,
without changing the NIC assignment determined above.

During dispatch, multiple experts selected by the same token may reside on
the same destination node. Sending their requests independently would transfer
the same token activation multiple times across the network. \Sys \textit{merges} such
requests into a shared inter-node transfer: the token crosses the network
once and reaches an initial receiver side GPU. It is then responsible for forwarding tokens over NVLink to the GPUs hosting experts that require them (Algorithm~\ref{alg:overlap}, lines~8--11).
As illustrated in Figure~\ref{fig:split_merge}(b), this replaces repeated
inter-node transfers with a single remote transfer and inexpensive node-local
forwarding. In addition, compatible small flows sharing the same communication
path can be coalesced into a larger transfer, reducing the number of
communication operations and amortizing per-transfer latency.

\textit{Merge} is applied differently during combine. Expert outputs are distinct and
therefore cannot be deduplicated. Instead, \Sys locally accumulates the
weighted expert outputs belonging to the same token into a node-local partial
sum. Once all local contributions are available, only the partial sum is
transferred back to the source node. This replaces multiple expert-level
return transfers with one node-level transfer while preserving every expert's
contribution.

\textit{Split} and \textit{merge} restructure inter-node communication into independently schedulable flows, but also introduce explicit dependencies between communication and expert computation. \Sys jointly schedules these flows and their dependent computation to minimize communication exposed on the critical path. The key is that dispatch and combine exhibit opposite dependencies: during dispatch, communication makes expert inputs ready for computation; during combine, computation makes node-local partial sums ready for communication. \Sys therefore makes computation follow communication during dispatch, but drive communication during combine.
Figure~\ref{fig:graph} illustrates the resulting schedule, and Algorithm~\ref{alg:overlap} summarizes the procedure; its \texttt{after ready} clauses express data dependencies, allowing communication and computation in separate loops to overlap.

\noindent\textbf{Dispatch follows communication.}
During dispatch, source tokens are immediately available, so \Sys first initiates the communication flows determined by the \textit{split}/\textit{merge} plan $\mathcal{P}$.
These flows progress in rounds across node pairs, while intra-node redistribution and forwarding realize their assigned physical paths, with local fan-out ordered cyclically across GPUs (Algorithm~\ref{alg:overlap}, lines~8--11). As soon as all inputs of a computation group $q$ arrive, the group becomes eligible for execution (lines~12--13); it does not wait for unrelated inputs of the same expert.

The computation order follows the planned delivery order rather than expert identity. \textit{Split} allows different input groups to become ready independently, while \textit{merge} makes a shared group ready after its inter-node transfer and subsequent local forwarding complete. \Sys prioritizes computation groups according to this communication order, allowing already-delivered groups to execute while later groups remain in transit.
In Figure~\ref{fig:graph}, Node~0 receives inputs from Node~1 in Round~1 and processes them on experts $E_0$--$E_3$ in Round~2. These experts thus execute a delivered input group while Node~0 receives the next group from Node~2.

\noindent\textbf{Combine drives communication.}
Combine reverses this dependency. Inter-node communication cannot begin until expert computation produces a complete node-local contribution for the destination node. In particular, sending an expert output immediately would forfeit the benefit of \textit{merge} and increase inter-node traffic; \Sys instead waits until all local expert contributions required by a token group are available, pre-reduces them into a node-local partial sum, and transfers the partial sum once (Algorithm~\ref{alg:overlap}, lines~22--25).

\Sys therefore schedules computation to make complete return groups ready as early as possible. Rather than processing tokens in an expert-major order, which can leave multiple destination groups blocked by their last unfinished experts, \Sys groups computation by the tokens' source nodes---their destinations $d$ during combine---and schedules the required expert computation across each group, denoted $\mathrm{returnGroup}[d]$ (Algorithm~\ref{alg:overlap}, lines~14--15). Once a group's contributions are complete, \Sys immediately performs local pre-reduction and initiates its return transfer, overlapping it with computation for subsequent groups.
In Round~1 of Figure~\ref{fig:graph}, completed contributions for Node~1 converge over NVLink on Node~0 and undergo pre-reduction ($\Sigma$) for return, while experts $E_0$--$E_3$ compute the Node~2 return group. In Round~2, that group's completed contributions enter the same aggregation and return path while these experts compute the local group.
At each source node, the arriving partial sums are reduced into the combined token outputs $y$ (Algorithm~\ref{alg:overlap}, line~26).

Together, the two schedules form complementary pipelines. During dispatch, \Sys consumes communication results as soon as they arrive, making communication drive computation; during combine, it reorders computation to release complete communication groups early, making computation drive communication. Across rounds, \Sys pipelines these dependencies so that inter-node transfers, intra-node movement, and expert computation progress concurrently whenever their dependencies permit. Expert movement introduces an additional readiness dependency, which we incorporate into the same schedule in \S\ref{sec:expert-swap}.

\begin{figure}[!t]
    \centering
    \includegraphics[width=1\columnwidth]{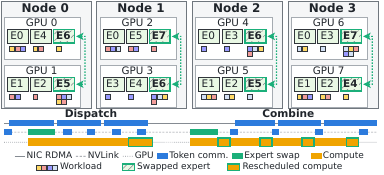}
    \vspace{-15pt}
    \caption{Intra-node expert swaps for dynamic GPU rebalancing. \Sys preserves node-level expert coverage while swapping replicas across local GPUs, and overlaps expert-weight movement with independent token communication and computation.}
    \label{fig:tps}
\end{figure}

\begin{figure*}[!ht]
    \centering
    \includegraphics[width=1\linewidth]{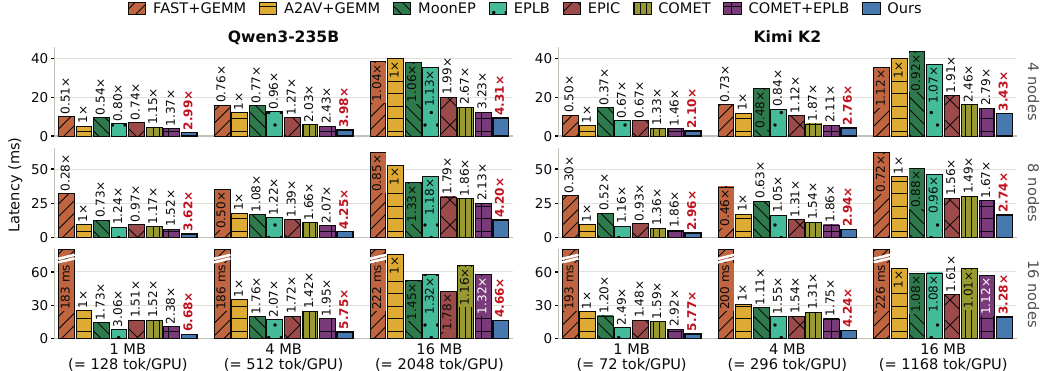}
    \vspace{-25pt}
    \caption{MoE layer latency on LiveCodeBench layer-5 traces~\cite{yu2026patternschaosforecastingdata} across node counts and token counts per GPU. Expert placements are initialized from frequencies in earlier decode steps, while evaluation workloads are sampled from later steps.}
    \label{fig:main-perf}
\end{figure*}

\subsection{Three Stream Scheduling}
\label{sec:expert-swap}

The preceding optimizations exploit a relatively stable expert placement to
reduce inter-node communication. However, as token demand changes across
batches, the historical placement may no longer satisfy the GPU compute-load
constraint. Globally remapping experts could restore balance, but would
introduce expensive inter-node weight transfers and disrupt the locality
established by placement and routing. \Sys instead performs lightweight
\emph{intra-node expert swaps}, which adjust GPU-level compute load while
preserving each expert's node-level coverage. Figure~\ref{fig:tps} illustrates
this process.

\noindent\textbf{Rebalancing under fixed node-level coverage.}
The initial placement spreads replicas of each expert across nodes
(\S\ref{sec:placement-routing}), providing multiple node-local opportunities
for subsequent adjustment. When the estimated GPU load exceeds the allowed
imbalance, \Sys searches for swaps between experts on overloaded and
underloaded GPUs within the same node. Let $\widehat L_j$ denote the estimated
compute load of GPU $j$, obtained from the current expert demand distributed
across its replicas. We select intra-node swaps to minimize the maximum GPU
load:
\begin{equation}
\label{eq:swap-objective}
    \min_{\mathcal{S}}\;\max_j \widehat L_j(\mathcal{S}),
\end{equation}
where $\mathcal{S}$ contains only swaps between GPUs in the same node.

A swap exchanges the physical GPU locations of two expert replicas while
preserving both GPU slot occupancy and the set of nodes hosting each expert.
It therefore changes neither expert replication nor node-level coverage, and
requires only intra-node weight movement. This restriction is deliberate:
rather than pursuing globally optimal compute balance at the cost of new
inter-node communication, \Sys performs best-effort GPU rebalancing within
the locality established by the initial placement. A swap is triggered only
when the current GPU-load imbalance exceeds the configured load constraint,
and is accepted only when it reduces the estimated maximum GPU load. The
selection policy and overhead are detailed in
\S\ref{sec:implementation}.

\noindent\textbf{Overlapping expert movement.}
Even intra-node swaps introduce weight movement that can interfere with token
communication and computation. Rather than treating swapping as a blocking
reconfiguration phase, \Sys incorporates expert-weight availability into the
dependency-aware schedule of \S\ref{sec:overlap-scheduling}. A computation
group becomes ready only when both its input tokens and the required expert
weights are available. \Sys therefore initiates weight transfers early and
executes independent expert groups while the moved weights are in transit.

As illustrated in Figure~\ref{fig:tps}, computation of unaffected experts
provides overlap windows for moving swapped experts. During dispatch, ready
experts consume arriving tokens while computation requiring a swapped expert
waits for its weights. During combine, \Sys preserves the destination-oriented
computation order from \S\ref{sec:overlap-scheduling}, but prioritizes
independent computation while the required weights are moving. Once a swapped
expert becomes available, its dependent computation proceeds without requiring
a global synchronization barrier. In this way, \Sys hides intra-node expert
movement whenever independent computation is available, while avoiding the
inter-node weight transfers of global expert remapping.

\section{Implementation}
\label{sec:implementation}

\Sys is implemented in C++ and CUDA~\cite{cuda_programming_guide} with Python bindings and uses CUTLASS~\cite{nvidia_cutlass} grouped GEMM kernels for expert computation. NVSHMEM~\cite{nvshmem} provides inter-node RDMA, while CUDA streams and NVSHMEM synchronization primitives coordinate transfers with expert computation.

\noindent\textbf{Routing.}
CUDA kernels translate expert demand and placement into locality-preferred token assignments that satisfy integer-rounded per-replica load bounds. A greedy water-filling procedure constructs these assignments by considering replicas on the source GPU first, followed by other GPUs within the same node and then remote nodes. Each allocation is limited by the receiving replica's remaining capacity while reserving sufficient unassigned demand to satisfy the other replicas' lower bounds. Starting from these feasible assignments, a greedy refinement consolidates each token's remote expert requests onto fewer nodes while preserving the same load bounds.

\noindent\textbf{Expert Swapping.}
Expert swapping uses a greedy intra-node search that pairs heavily and lightly loaded GPUs, accepting exchanges only when they reduce the pair's maximum estimated load.
Together, routing, swap planning, and associated metadata exchange and construction account for approximately 5\% of iteration latency (\S\ref{sec:case_study}).

\noindent\textbf{Transport.}
Dispatch implements union forwarding by having each receiving node GPU forward complete received tokens to all GPUs within the destination node. Expert GEMMs select their required rows from these tokens through gather indices, eliminating recipient-specific forwarding-index construction and repacking. These transfers use NVSHMEM symmetric staging buffers to hold intermediate payloads for dispatch redistribution and combine aggregation. The buffers are allocated during initialization and reused across iterations.

%% file: 05_Evaluation.tex
\section{Evaluation}
\subsection{Experimental Setup}
\label{sec:experimental-setup}

\noindent\textbf{Testbed.}
Each node contains four 40\,GB memory NVIDIA A100 GPUs; \S\ref{sec:e2e-performance} uses 80\,GB variants to accomodate for KV caches and other serving state.
Intra-node GPU pairs are connected by third-generation NVLink, providing 100\,GB/s bandwidth per direction~\cite{nvidia2020a100architecture}.
In \S\ref{sec:scalability}, we additionally use four 96\,GB NVIDIA H100 GPUs per node with the same GPU/NIC topology, fourth-generation NVLink providing 150\,GB/s per direction between GPU pairs~\cite{nvidia2022h100architecture}.
Both configurations bind one HPE Slingshot~11 NIC per GPU providing 25\,GB/s of inter-node bandwidth~\cite{hpe_slingshot_quickspecs}. Our software uses
CUDA~12.4, PyTorch~2.6.0, and NVSHMEM~3.2.5 with the libfabric/CXI transport~\cite{cuda_programming_guide,paszke2019pytorchimperativestylehighperformance, nvshmem}.

\noindent\textbf{Baselines.}
We compare our system with various communication and expert replication \& placement baselines, representative of different approaches to MoE layer latency optimizations:
A2AV+GEMM (implemented using NVSHMEM~\cite{nvshmem}), FAST+GEMM~\cite{lei2026fast}, COMET~\cite{zhang2025comet}, EPLB~\cite{deepseek2025eplb}, EPIC~\cite{cao2026epic} and MoonEP~\cite{moonshot2026moonep}. We additionally evaluate COMET+EPLB, which combines EPLB's expert replication and placement with COMET's communication--computation overlap while retaining COMET's dense allgather dispatch.
Expert replication baselines and \Sys use two extra expert slots per GPU
rank $(R=2)$.

\noindent\textbf{Workloads.}
We use publicly released MoE layer expert traces datasets from prior work~\cite{yu2026patternschaosforecastingdata} for Kimi-K2-Thinking (K2)~\cite{kimiteam2026kimik2openagentic} and Qwen3-235B-A22B (Qwen)\cite{yang2025qwen3technicalreport}. Table~\ref{tab:eval-models} lists their MoE dimensions; K2's greater amount of routed experts challenges expert replication baselines, and its larger expert FFN dimension increases the computation cost per-token and communication cost of moving experts. Our experiments also fix parameters to use BF16.
Within the selected dataset, we sample complete per-token top-$k$ selections with replacement to reach the target workload size, measured as BF16 input token activation volume per GPU \emph{before} top-$k$ replication.

\begin{table}[!ht]
  \centering
  \small
  \setlength{\tabcolsep}{3pt}
  \renewcommand{\arraystretch}{1.1}
  \begin{tabular*}{\columnwidth}{@{\extracolsep{\fill}}lrrrr@{}}
    \toprule
    Model & \shortstack{Routed\\experts} & \shortstack{Token\\dim.}
          & \shortstack{Expert\\FFN dim.} & Top-$k$ \\
    \midrule
    Kimi-K2-Thinking & 384 & 7,168 & 2,048 & 8 \\
    Qwen3-235B-A22B & 128 & 4,096 & 1,536 & 8 \\
    \bottomrule
  \end{tabular*}
  \vspace{-12pt}
  \caption{Model configurations.}
  \label{tab:eval-models}
\end{table}

\noindent\textbf{Measurement.}
We measure the maximum per-GPU elapsed time of all GPUs engaged in a single MoE layer as the iteration latency. The timed window includes all metadata exchange, placement update planning, token routing planning, and token communication and computation in both dispatch and combine. We run and exclude one-time initialization and warmup iterations. 

\subsection{MoE Performance}
\label{sec:moe-performance}
Figure~\ref{fig:main-perf} shows that \Sys{} achieves $2.10$--$6.68\times$ speedup over A2AV+GEMM and up to $2.78\times$ over the fastest competing baseline. We examine these gains through the roles of placement and locality-preferred routing across node counts, and of traffic reduction and overlap across workload sizes. At sixteen nodes, we contrast the workload-size extremes to examine the value of avoiding remote requests and overlapping the remaining transfers.

At 16\,MB, \Sys{}'s advantage over COMET+EPLB grows with node count. Between four and sixteen nodes, our speedup rises from $1.34\times$ to $3.53\times$ for Qwen and from $1.23\times$ to $2.94\times$ for K2. This growing advantage is consistent with reducing the remote traffic that COMET+EPLB's dense allgather leaves to overlap, even after expert replication. At four nodes and 1\,MB, Zepp already outperforms COMET+EPLB by $2.18\times$ for Qwen and $1.44\times$ for K2. Here, placement and locality-preferred routing exploit available local replicas to serve requests within their source nodes (\S\ref{sec:placement-routing}), whereas dense allgather still sends tokens to remote nodes.

Larger workload sizes increase the potential savings from traffic reduction and the work available for overlap. At four nodes, increasing the workload size from 1 to 16\,MB raises our speedup over A2AV+GEMM from $2.99\times$ to $4.31\times$ for Qwen and from $2.10\times$ to $3.43\times$ for K2. Our intra-node expert swaps also avoid MoonEP's inter-node weight transfers, whose cost is especially difficult to amortize at small workload sizes. EPLB's static placement similarly avoids recurring weight movement, but our advantage over it supports optimizing the token communication that remains after compute balancing. For this remaining traffic, EPIC's improving performance relative to EPLB at larger workload sizes reinforces the value of merge, which \Sys{} combines with overlapping the remaining transfers.

At sixteen nodes, avoiding remote communication remains valuable at small workload sizes even without complete expert coverage across nodes. At 1\,MB, \Sys{} achieves $6.68\times$ and $5.77\times$ speedups over A2AV+GEMM for Qwen and K2, respectively. Locality-preferred routing exploits available local replicas under GPU load constraints (\S\ref{sec:placement-routing}), reducing remote requests when small payloads offer little opportunity to amortize inter-node latency. FAST's high sixteen-node latency further illustrates the importance of controlling communication coordination overhead, which can accumulate as node count grows. At 16\,MB, Zepp achieves $4.66\times$ and $3.28\times$ speedups: larger transfers increase the savings from traffic reduction, while longer computation intervals provide more opportunity to hide the remaining communication. Across all configurations, these optimizations achieve a geometric mean speedup of $1.86\times$ over the fastest competing baseline in each configuration.

\begin{figure}[!t]
    \centering
    \includegraphics[width=1\columnwidth]{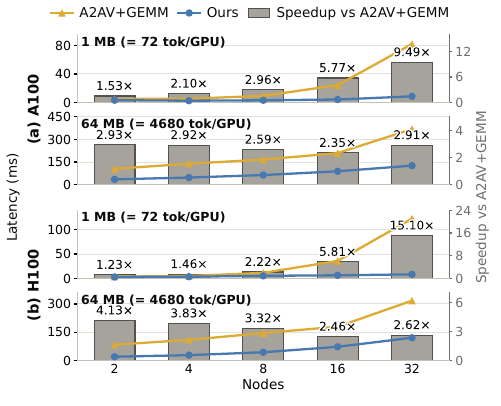}
    \vspace{-30pt}
    \caption{Weak scaling of K2 MoE layer latency on LiveCodeBench traces across 2--32 nodes (8--128 GPUs) on (a) A100 and (b) H100 clusters. Speedups are relative to A2AV+GEMM.}
    \label{fig:scalability}
\end{figure}

\subsection{Weak Scalability}
\label{sec:scalability}

We evaluate weak scalability on A100 and H100 (Figure~\ref{fig:scalability}) by increasing the cluster from 2 to 32 nodes while holding input token volume per GPU fixed, so that the aggregate workload grows with node count. The 1 and 64\,MB workload sizes represent small and large batches relevant to latency-oriented and throughput-oriented serving, respectively.

At two nodes on A100, \Sys{} benefits substantially more from the larger workload, achieving $2.93\times$ speedup at 64\,MB compared with $1.53\times$ at 1\,MB.
H100 exhibits the same contrast, with $4.13\times$ speedup at 64\,MB compared with $1.23\times$ at 1\,MB.
At 1\,MB, the small communication volume limits the absolute savings from reducing traffic, while short computation intervals restrict overlap. Locality-preferred routing also has limited flexibility, with only 16 extra expert slots for 384 experts. Increasing the workload size to 64\,MB lengthens the computation and transfer intervals, providing more opportunity to overlap them and helping explain the larger speedup.

As the cluster grows, the small-workload advantage strengthens while the large-workload advantage persists.
At 1\,MB, speedup reaches $9.49\times$ on A100 and $15.10\times$ on H100 at thirty-two nodes. Over this scaling range, the extra expert slots expand from 16 to 256, giving placement and locality-preferred routing more flexibility to avoid remote requests. 
Avoiding these requests is particularly valuable for small payloads, whose transfers are sensitive to initiation and synchronization costs.
At 64\,MB, speedup stays within $2.35$--$2.93\times$ on A100 and $2.46$--$4.13\times$ on H100 across all evaluated node counts, showing that the combined benefits of communication reduction and overlap persist as the aggregate workload increases.
The growing small-workload advantage and sustained large-workload gains on H100 also demonstrate that \Sys{}'s optimizations remain effective on newer GPU hardware.

\begin{figure}[!t]
    \centering
    \includegraphics[width=1\columnwidth]{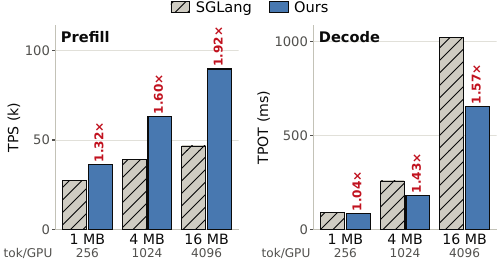}
    \vspace{-15pt}
    \caption{End-to-end performance of \Sys integrated into SGLang~\cite{sglang}: Tokens Per Second (TPS, higher is better) throughput during prefill and Time Per Output Token (TPOT, lower is better) latency during decode. Evaluated using Qwen3-30B which shares same expert count and top-$k$ as Qwen3-235B with half the hidden dimension, doubling token counts per GPU compared to figure \ref{fig:main-perf}.}
    \label{fig:e2e}
\end{figure}

\subsection{End-to-End Performance}
\label{sec:e2e-performance}
Next, we evaluate whether \Sys{}'s MoE-layer gains translate to end-to-end performance by integrating it into SGLang~\cite{sglang} while retaining SGLang's gating and expert selection.
At each workload size, both configurations use data-parallel attention and matched KV-cache sizes, with the SGLang baseline executing fused all-gather, expert computation, and reduce-scatter sequentially without overlap.
We use Qwen3-30B instead of Qwen3-235B to leave sufficient memory for KV caches at larger batch sizes. Placement is now calculated from a subset of LiveCodeBench prompts, while evaluation uses unseen inputs from the same input dataset.

Figure~\ref{fig:e2e} shows that \Sys{} improves both prefill throughput and decode latency, with gains increasing as input token volume grows.
From 1 to 16\,MB per GPU, prefill throughput gains increase from $1.32\times$ to $1.92\times$, while decode speedups increase from $1.04\times$ to $1.57\times$.
This widening advantage shows that MoE layer optimizations remains consequential even when attention and other model operations contribute to execution time.
At smaller batch sizes, more variable expert demand triggers more frequent intra-node swaps to correct deviations from historical placement estimates, while less available computation limits the opportunity to hide their cost, consistent with the modest gains and near-parity decode performance at 1\,MB.

The reported end-to-end gains are achieved while making these runtime decisions at each MoE layer without per-layer host synchronization. Planning and routing execute on the GPU, with fused operations reducing host-issued work. GPU dependencies preserve execution order, allowing the host to continue enqueueing subsequent layers while these decisions are resolved on the GPU device.

\begin{figure}[!t]
    \centering
    \includegraphics[width=1\columnwidth]{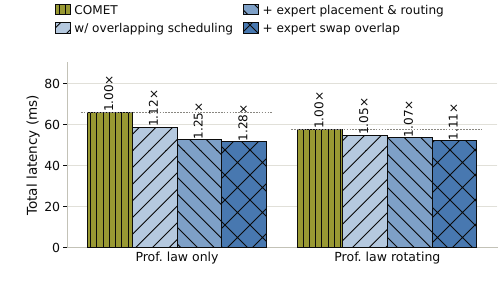}
    \vspace{-15pt}
    \caption{Component ablation of \Sys on K2, 4 nodes, with 64 MB of input tokens per GPU.}
    \label{fig:ablation}
\end{figure}

\begin{figure*}[!t]
    \centering
    \includegraphics[width=1\linewidth]{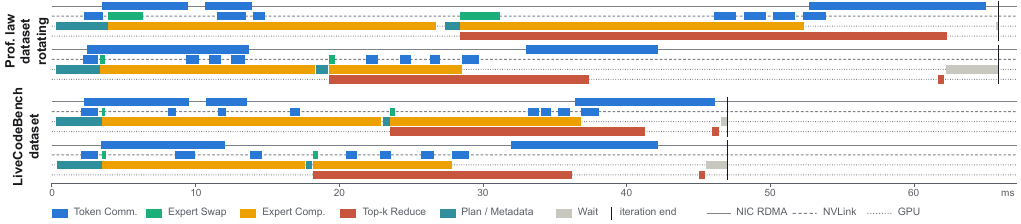}
    \vspace{-25pt}
    \caption{Resource usage timelines under K2 model architecture on 4 nodes, 64\,MB of input tokens per GPU. Under each dataset, the GPUs with the longest and shortest total GEMM durations are plotted.}
    \label{fig:case_study}
\end{figure*}

\subsection{Ablation}
We ablate \Sys{}'s components under two separate demand datasets on four nodes (Figure~\ref{fig:ablation}), where COMET provides a competitive overlap baseline. Prof.~law's greater imbalance shown in Figure \ref{fig:intro} exposes opportunities for expert swaps that remain limited under the LiveCodeBench setting. In \emph{Prof.~law dataset only}, we initialize placement from an equal mixture of eight datasets but serve only Prof.~law, modeling concentrated demand for a previously observed dataset. In \emph{Prof.~law dataset rotating}, we exclude Prof.~law from the placement history and cycle through all eight datasets for four iterations each, modeling occasional unexpected demand.

With Prof.~law represented in the placement history, overlapping scheduling and placement/routing provide most of the performance gain. Overlapping scheduling alone achieves a $1.12\times$ speedup over COMET without expert replication, demonstrating the benefit of reducing and redistributing traffic on top of compute communication overlap (\S\ref{sec:overlap-scheduling}). Adding replication, placement, and locality-preferred routing raises speedup to $1.25\times$, as replicas provide opportunities to balance computation and serve requests locally (\S\ref{sec:placement-routing}). Overlapped expert swaps further increase speedup to $1.28\times$, a $1.9\%$ latency reduction. This modest increment is consistent with the historical mixture already accounting for Prof.~law, leaving swaps to correct residual imbalance when demand concentrates on that dataset.

Under rotating demand, overlapped swaps contribute more despite smaller overall gains. Overlapping scheduling achieves a $1.05\times$ speedup, and adding placement/routing raises it to $1.07\times$; averaging across seven less imbalanced datasets dilutes the benefits of these stages. Overlapped swaps nevertheless increase speedup to $1.11\times$, a further $3.0\%$ latency reduction compared with $1.9\%$ under Prof.~law-only demand. Because Prof.~law is absent from the placement history, its arrival exposes compute imbalance that intra-node swaps can correct without inter-node weight transfers (\S\ref{sec:expert-swap}).

\subsection{Case Study}
\label{sec:case_study}
Figure~\ref{fig:case_study} shows how \Sys{} overlaps token communication and intra-node expert swaps with computation under different demand conditions. We contrast a Prof.~law iteration from the rotating workload, whose placement history excludes Prof.~law, with LiveCodeBench served using history from the same workload. Each profiled iteration starts from the static placement to expose expert swaps.

Under Prof.~law, \Sys{} shows intra-node expert swaps on NVLink and overlaps these transfers with expert computation. The selected ranks accumulate 24.2 and 46.7\,ms of expert GEMM execution, illustrating the uneven execution associated with this demand shift. To help alleviate the skew, intra-node swaps can be seen on the timeline occupying the NVLink resource, keeping expert-weight traffic off inter-node links. The resulting weight-copy bursts overlap both GEMMs, while inter-node dispatch overlaps the first GEMM: scheduling ready expert work ahead of swapped-expert work allows token delivery, weight movement, and computation to proceed concurrently. Expert outputs then enter combine as they become ready, although residual imbalance remains visible in late completion, consistent with the intra-node scope of rebalancing.

Under LiveCodeBench, \Sys{} combines more balanced expert execution with overlap across dispatch, expert movement, and combine. The selected ranks accumulate 23.8 and 32.8\,ms of expert GEMM execution, consistent with placement and routing matching current demand. These computation windows accommodate dispatch alongside the first GEMM, allowing ready token groups to enter computation while remaining transfers proceed. Alongside token delivery, the much smaller intra-node expert swaps overlap their corresponding GEMMs and inter-node transfers. As expert outputs become ready, \Sys{} overlaps their pre-reduction for merge and inter-node return transfers with the second GEMM; late-arriving contributions and their final top-$k$ reduction form the remaining tail.

%% file: 06_Conclusion.tex
\section{Related Work}
\label{sec:related-work}

\noindent\textbf{Communication optimizations.}
Communication optimizations improve data movement and network utilization. FAST~\cite{lei2026fast} rebalances skewed all-to-all traffic over fast intra-node links and schedules inter-node transfers to avoid incast. FuseLink~\cite{ren2025fuselink} redirects traffic through intra-node connections to exploit otherwise idle NICs. SwiftEP~\cite{li2026swiftep} reduces MoE communication overhead through buffer fusion and TMA offloading. We introduce \textit{split} and \textit{merge} as graph transformations that jointly restructure MoE token transfers and their synchronization dependencies across intra- and inter-node links.

\noindent\textbf{Computation load balancing.}
Expert replication and placement distribute computation under skewed routing. EPLB~\cite{deepseek2025eplb} replicates popular experts and packs them onto GPUs according to estimated loads; its hierarchical policy also co-locates expert groups to reduce inter-node traffic. EPIC~\cite{cao2026epic} uses communication-aware expert placement and replication, pipelines communication and computation across expert groups, and eliminates duplicate transfers. CRAFT~\cite{zhao2026craft} allocates replicas across layers according to their expected benefit under a memory budget. MoonEP~\cite{moonshot2026moonep} uses current routing decisions to plan redundant experts and prefetch their weights for balanced execution. Our placement spreads hot-expert replicas across nodes and couples them with locality-preferred token routing under per-GPU load constraints.

\noindent\textbf{Computation--communication overlap.}
Overlap exposes computation before all communication. NVSHMEM~\cite{nvshmem} provides fine-grained, GPU-initiated one-sided communication and synchronization within CUDA kernels, enabling persistent kernels to interleave communication with computation. COMET~\cite{zhang2025comet} resolves fine-grained producer--consumer dependencies to overlap communication with tiled GEMMs. FlashMoE~\cite{aimuyo2025flashmoe} fuses dispatch, expert computation, and combine into a persistent GPU kernel with tile-level scheduling and device-initiated transfers.  StreamEP~\cite{liang2026streamep} removes collective barriers during MoE decoding through per-layer queues and asynchronous rebatching. Our scheduler coordinates expert computation with the dependencies produced by \textit{split} and \textit{merge}.

\noindent\textbf{Collective communication libraries.}
Collective libraries provide reusable communication interfaces and execution mechanisms. NCCL~\cite{nccl} provides topology-aware collective and point-to-point communication across GPUs and nodes. MSCCL++~\cite{hwang2026mscclpp} exposes low-level communication primitives, a DSL, and a collective API for portable GPU communication. DeepEP~\cite{deepseek2025deepep} specializes dispatch and combine kernels for expert parallelism. Related scheduling work includes SyCCL~\cite{cao2025syccl}, which exploits collective and topology symmetry, and HeteCCL~\cite{hei2026heteccl}, which synthesizes schedules for heterogeneous clusters. ForestColl~\cite{zhao2026forestcoll} constructs throughput-optimal broadcast and aggregation schedules, while ResCCL~\cite{liu2025resccl} optimizes primitive execution and GPU resource allocation. PReCCL~\cite{chen2026preccl} reallocates collective work using network telemetry, and Theseus~\cite{ding2026theseus} supports runtime replacement of collective schedules.

\section{Conclusion}
We presented \Sys{}, a distributed MoE serving system that directly optimizes bottleneck inter-node communication. By treating GPU and NIC balance as resource constraints, \Sys{} addresses the communication overhead introduced by pursuing balance independently. Under these constraints, it combines expert replication, placement, and routing with communication reduction and communication--computation overlap, while adapting to changing demand. Evaluation against 7 baselines on 4--16 nodes shows up a geometric mean speedup of $1.86\times$ over the fastest competing baseline, while weak-scaling experiments demonstrate sustained gains on up to 32 nodes.